\documentstyle[aps,graphics,prl,twocolumn]{revtex} 

\begin{document}
\draft

\title{Anisotropy of Quasiparticle Lifetimes and the Role of Disorder for Ultrafast Electron Dynamics in Graphite}

\author{Gunnar Moos, Cornelius Gahl$^\dag$, Roman Fasel$^\ast$, Martin Wolf$^\dag$ and Tobias Hertel}
\address{Fritz-Haber-Institut der Max-Planck-Gesellschaft,
Faradayweg 4--6, 14195 Berlin, Germany\\$^\dag $Freie
Universit\"{a}t Berlin, Institut f\"{u}r
Experimentalphysik, Arnimallee 14, 14195 Berlin,
Germany
\\$^\ast$EMPA D\"{u}bendorf, \"{U}berlandstr. 129, 8600 D\"{u}bendorf, Switzerland}

\date{\today}
\maketitle

\begin{abstract}

Femtosecond time-resolved photoemission of photoexcited electrons
in highly oriented pyrolytic graphite (HOPG) provides strong
evidence for anisotropies of quasiparticle (QP) lifetimes.
Indicative of such anisotropies is a pronounced anomaly in the
energy dependence of QP lifetimes between 1.1\,eV and 1.5\,eV \---
the vicinity of a saddle point in the graphite bandstructure. This
is supported by recent \emph{ab initio} calculations and a
comparison with experiments on defect-enriched HOPG which reveal
that disorder, {\it e.g.} defects or phonons, increases electron
energy relaxation rates.

\end{abstract}


\narrowtext

Studies of the ultrafast dynamics of electronic excitations in
solids have challenged experiment and theory for years. Advances
in ultrafast laser technology and the latest theoretical
developments allow us to investigate electron dynamics in growing
detail and further our understanding of fundamental scattering
processes in solids at the femtosecond time scale
\cite{Pet97,Ech00}.
\par
For an interacting 3D-electron gas the standard theory
of \emph{e-e} scattering \--- Landau's theory of
Fermi-Liquids \--- predicts a quadratic dependence of
scattering rates on the quasiparticle (QP) energy
$(E-E_F)$ \cite{Pin89}. In a periodic potential,
however, the electronic states are modified with
respect to those of a free electron gas and form Bloch
states which may result in different QP lifetimes at
the same energy if distinct $k$--states within the
Brillouin zone are compared. Such anisotropies were
indeed predicted by recent {\it ab initio} self-energy
calculations for Al and Cu, which find strong
variations of QP lifetimes for electrons in different
bands \cite{Sch99,Cam99}. Experimental verification of
these predictions, however, remains truly challenging.
Time-resolved photoemission from different copper
surfaces has revealed some dependence of the measured
electron dynamics on the crystallographic surface
orientation, but the observed effects could not be
clearly attributed to anisotropies of QP lifetimes
\cite{Oga97}. Similar experiments on aluminum
\cite{Bau98} have likewise not been able to identify
band structure effects predicted theoretically
\cite{Sch99}.

\par
Graphite, a semi-metal with layered structure, is
expected to be an ideal candidate if band structure
effects are to be observed experimentally. The
strongly anisotropic band-structure of graphite is
expected to furnish electron scattering processes with
similar anisotropies, leading to anomalous QP
lifetimes. The special topology of the graphite
bandstructure has indeed been used to explain apparent
deviations of the electron dynamics observed on a
ceasiated HOPG surface \cite{Xu96} from the standard
predictions for a 3D-electron gas \cite{Gon96}.
However, the experiments provided no evidence for
anomalies that might be associated with anisotropic QP
lifetimes. Furthermore, the overlap of the energy
range probed experimentally in Ref.\,\cite{Xu96} and
the range of validity of the approximations used in
Ref.\,\cite{Gon96} is small and calls for more
extended and more detailed theoretical as well as
experimental work.
\par
Here we present a time-resolved photoemission study of the
electron dynamics in HOPG, with particular focus on its energy
dependence and the influence of disorder on the relaxation rates.
The most striking observation \--- a pronounced anomaly in the
energy dependence of the electron dynamics between 1.1\,eV and
1.5\,eV \--- can be associated with electrons near a saddle point
in the graphite band structure at the $M$-point of the Brillouin
zone. This, and a comparison with experiments on defect enriched
HOPG provide strong evidence for anisotropies of QP lifetimes in
graphite which have also been predicted by a recent \emph{ab
initio} calculation \cite{Spa01}.

The HOPG sample \cite{advcer} was attached to a tantalum disk
which could be temperature controlled from 25\,K to 1200\,K. It
was mounted in an UHV chamber with a base pressure of $2\times
10^{-10}$\,mbar. Experiments were performed at room temperature,
unless mentioned otherwise. The HOPG sample was cleaved directly
before being transferred into UHV and heated repetitively to
900\,K. The defect enriched HOPG surface was produced by Ar-ion
sputtering (0.5\,keV, 7\,$\mu A$, 300\,s).

\par
Photoelectron spectra are obtained by means of the
time-of-flight technique with an energy resolution of
10\,meV. For time resolved pump-probe measurements a
visible pump and frequency doubled probe pulse of
typically 85\,fs duration are focused on a spot of
50\,$\mu$m diameter on the sample. The photon energy
of the probe pulse was chosen to exceed the work
function of the sample ($e\Phi=4.50$\,eV $\pm
0.05$\,eV) by typically 0.15\,eV in order to probe not
only low energy electron but also hole dynamics. A
more detailed description of the experiment can be
found elsewhere \cite{Kno98}.

\par
We start our discussion with a brief overview of
processes contributing to the dynamics of photoexcited
carriers. After optical excitation the energy absorbed
by the electrons is redistributed via
electron-electron scattering (\emph{e-e}),
electron-phonon scattering (\emph{e-ph}) and transport
from the surface into the bulk. In graphite the
transport of electrons out of the photoemission
detection volume into the bulk is hampered due to the
comparatively weak interlayer coupling and is further
suppressed by frequent stacking faults in HOPG
\cite{Ono76}. Therefore, the dynamics observed in the
experiments are expected to be dominated by \emph{e-e}
and \emph{e-ph} scattering processes with \emph{e-ph}
scattering becoming increasingly important closer to
the Fermi level.

\begin{figure}
\center

\includegraphics{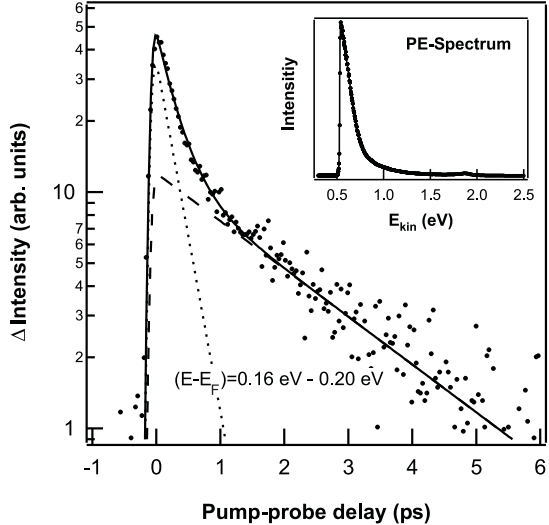}

\vspace{0.5cm}

\caption{Cross correlation trace: change of the photoemission
intensity within a certain energy interval as a function of the
pump-probe time-delay. The pump photon energy was 2.3\,eV. The
solid line is the fit to a bi-exponential decay. Fast and slow
components are indicated by the dotted and dashed lines
respectively. The inset shows a 2PPE-spectrum for simultaneous
pump-probe excitation.}\label{fig1}
\end{figure}

\par

A typical photoelectron spectrum for simultaneous pump
and probe excitation is shown in the inset of
Fig.\,\ref{fig1}. The electron dynamics are obtained
by plotting the change $\Delta I$ of the photoemission
intensity, resulting from excitation by the visible
pump pulse, as a function of the pump-probe time-delay
(see Fig.\,\ref{fig1}). These cross correlation traces
(XCs) can clearly be characterized using a
bi-exponential decay with a fast component decaying on
the sub-picosecond time-scale and a slower component
decaying on the ps time-scale.

A detailed analysis of photoelectron spectra obtained
at different pump-probe time-delays shows that the
electronic system approaches an internal equilibrium
with a characteristic time constant of
$(250\pm50)\,$fs. This is estimated using the
deviation of photoelectron spectra from the best fit
to a Fermi-Dirac distribution \cite{Her00a,Moo01}.
This process \--- referred to as internal
thermalization \--- is associated with the initial
fast decay seen in XC-traces and results in an
increase of the electron gas temperature by 400\,K.
This is in reasonably good agreement with the
calculated temperature rise using the electronic heat
capacity of graphite and the estimated laser power
deposited within the optical penetration depth. The
slow component of the XCs can be associated with
thermalization of the electronic system with the
lattice due to {\emph e-ph} interaction
\cite{Moo01,Her00,Sei90}.

\par
\begin{figure}
\center

\includegraphics{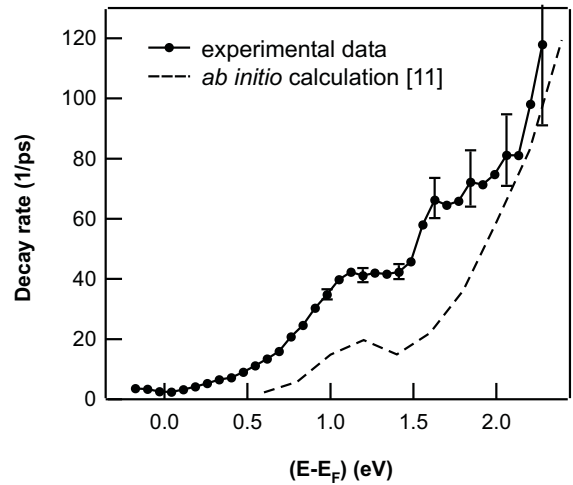}

\vspace{0.5cm}

\caption{Energy dependence of decay rates on the pristine HOPG
surface. The anomaly near 1.5\,eV is indicative of anisotropic QP
lifetimes. This is supported by a recent {\it ab inito}
calculation of QP relaxation rates for graphite which shows a
similar anomaly in the same energy region [11].  } \label{fig2}
\end{figure}

 In the following we focus on the initial
fast decay, associated with the internal
thermalization of the electronic system. The
corresponding decay rates \--- as obtained from the
fast component of the bi-exponential fit \--- are
plotted in Fig.\,\ref{fig2} as a function of the
intermediate state energy
$(E-E_F)=E_{kin}+e\Phi-h\nu_{probe}$. The data shown
in Fig.\,\ref{fig2} extend to energies $\sim0.15\,$eV
below $E_F$, and present, to the best of our
knowledge, the first 2PPE measurements of electron
{\it and} hole dynamics close to $E_F$. Note, that
lifetimes of holes in the copper {\emph d}-bands have
been deduced indirectly from hot electron lifetimes by
Petek and coworkers \cite{Pet00}.

Obviously the energy dependence of the decay rates
cannot be described by a $(E-E_F)^n$-dependence with
$n=2$ as expected for a Fermi Liquid or $n=1$ as
previously predicted for a simplified bandstructure
model of graphite \cite{Gon96}. The most striking
deviation from a simple power law is the anomaly
between 1.1\,eV and 1.5\,eV. In this energy range we
observe a plateau-like region with no significant
increase of the relaxation rates. The error bars in
Fig.\,2 reflect statistical errors obtained from the
bi-exponential fit after averaging over 13
 experimental runs. Systematic errors due to day-to-day
variations are somewhat larger.

\begin{figure}
\center

\includegraphics{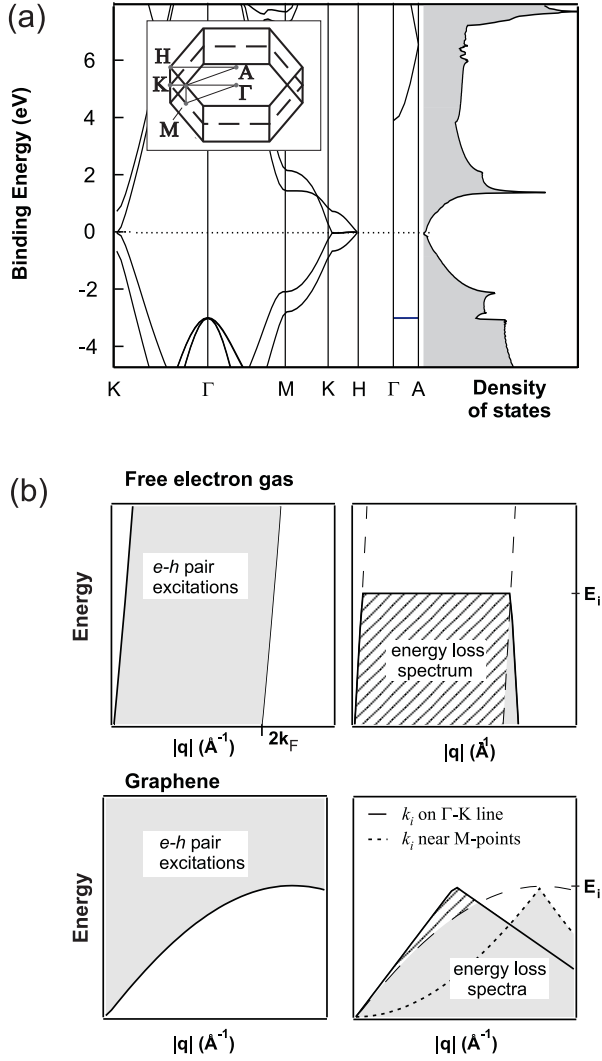}

\vspace{0.5cm}

\caption{(a) Graphite bandstructure along high symmetry lines and
density of states. (b) {\emph e-h} pair spectrum for a free
electron gas and a graphene sheet (left panels) as well as
energy\--loss {\it vs.} momentum\--change for electrons decaying
from an intermediate state with $E_i = 1.5\,$eV (right panels).
The overlap of \emph{e-h} pair excitation- with energy
loss-spectra is given by the hatched area.}\label{fig3}
\end{figure}

\par
The most obvious feature in the graphite bandstructure with which
this plateau may be associated is a saddle point of the
$\pi^\ast$-bands at the $M$-point of the graphite Brillouin zone
(see Fig.\,\ref{fig3}a)) and the resulting van Hove singularity at
1.5\,eV in the DOS. Electrons excited to this region of the
band-structure can only decay by a transfer of comparatively large
momentum to secondary excitations (\emph{e-h} pairs), in contrast
to electrons decaying from other iso-energetic regions of the
bandstructure. This is illustrated schematically in the lower
right panel of Fig.\,\ref{fig3}b) for a simplified band structure
model of a graphene sheet \cite{Wal47}. The {\emph e-h} pair
excitation spectrum of graphene on the other hand (lower left
panel) has no overlap with the energy\--loss and associated
momentum\--change for electrons decaying from the vicinity of the
$M$-point. In contrast, a much smaller mismatch exists for
iso-energetic electrons along the $\Gamma-K$ line. We speculate
that this gives rise to strongly anisotropic QP lifetimes in
graphite and ultimately leads to the observed anomaly in the decay
rates. Note, that no comparable anisotropies exist for a free
electron gas (upper panels in Fig\,\ref{fig3}b)) or for lower
energy excitations in graphene.

\par
Recent \emph{ab initio} calculations of QP lifetimes in graphite
by Spataru {\it et al.} \cite{Spa01} show a similar anomaly around
1.5\,eV. The authors find pronounced anisotropies in the QP
lifetimes with a distinct anomaly at 1.5\,eV arising from
electrons in the $\pi^\ast$-band around the $M$-point of the
Brillouin zone. If averaged over the entire Brillouin zone \---
mimicking the averaging process in 2PPE due to scattering in the
intermediate and final states \--- the calculated decay rates are
reproduced by the dashed line in Fig.\,2. The qualitative trends
and features of both, experimental and theoretical decay rates are
evidently the same.

\par
The calculations by Spataru {\it et al.} strongly support, that
the experimentally observed plateau around 1.5\,eV is a direct
consequence of anisotropic QP lifetimes in graphite. A
quantitative comparison of theoretical and experimental results,
however, appears to be difficult. This is most likely because
theory does not account for finite temperatures, unavoidable
lattice defects, and possible corrections from diffusive or
ballistic transport of excited carriers out of the detection
volume; effects which are expected to lead to higher decay rates
as observed experimentally.

\par

The influence of disorder, \emph{e.g.} defects or phonons, on the
electron dynamics is expected to depend strongly on anisotropies
of QP lifetimes. This is best understood if one considers that
lattice distortions break the translational symmetry of the
crystal lattice and, thereby, lead to a coupling of different
electronic states. If the states coupled in this manner have
different lifetimes this should ultimately change the global
electron dynamics. Matthiesens rule \--- which states that the
rates of competing decay channels simply add up \--- would imply
that the global dynamics is increasingly dominated by short-lived
states once the coupling among states in the Brillouin zone is
turned on.

\par
\begin{figure}
\center

\includegraphics{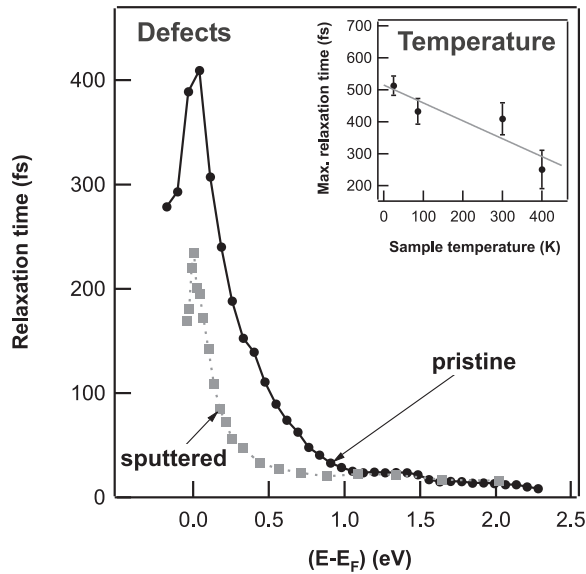}

\vspace{0.5cm}

\caption{Influence of defects and lattice temperature on the
electron dynamics.}\label{fig4}
\end{figure}

To test this hypothesis we introduced lattice defects
by Ar-ion sputtering. In Fig.\,\ref{fig4} we plot the
relaxation times of the electron population \--- the
inverse of the rate \--- as a function of intermediate
state energy. Qualitatively we find that the decay of
the electronic population is accelerated significantly
irrespective of the electron energy.

\par
In agreement with the arguments presented above, these
results provide further evidence for anisotropies in
the QP lifetimes in graphite. Furthermore, they extend
a recent study on the influence of defects on the
electron dynamics in surface states \cite{Wei99} to
the dynamics in bulk states.

\par
Alternatively, coupling between electronic states can
be introduced by heating of the sample, \emph{i.e.} by
an increase of dynamic lattice distortions. This is
most clearly illustrated by plotting the decay-time
found at the peak of relaxation time curves like the
one in Fig.\,\ref{fig4}, as a function of lattice
temperature (see inset of Fig.\,\ref{fig4}). Very much
like static defects, lattice vibrations can lead to a
coupling of electronic states and reduce the measured
relaxation times. We note, however, that the trend
seen in Fig.\,\ref{fig4} may be a superposition of
different effects with additional contributions from
the temperature dependence of \emph{e-e} scattering
rates and enhanced inelastic \emph{e-ph} scattering.

\par

These results may also shed new light on the apparent
discrepancies between electron dynamics measured by
different groups in graphite. Xu \emph{et al.}, for
example, have observed somewhat shorter relaxation
times on a cesiated graphite surface \cite{Xu96} whose
absolute values lie in-between our results for the
pristine and the sputtered surface. Furthermore, as in
our experiments on the defect enriched surface,
neither Xu \emph{et al.} nor Ertel \emph{et al.}
\cite{Ert99} find any evidence for a plateau around
1.5\,eV \. This may indicate that surface preparation
and defect density plays a more important role for
measurements of bulk electron dynamics than previously
assumed.

\par
In summary, we have investigated the electron dynamics in HOPG
resulting from ultrafast optical excitation. The observed anomaly
in the energy dependence of the electron dynamics can be linked to
strongly anisotropic lifetimes of electrons excited to about
1.5\,eV above the Fermi level. Electrons excited to a saddle point
in the graphite bandstructure at the same energy are expected to
be very long lived due to a mismatch of their energy loss spectrum
and the {\emph e-h} pair excitations available for scattering
processes. Disorder is found to accelerate global electron
dynamics, most likely due to the coupling of long- to short-lived
states. This provides further evidence for the anisotropy of QP
lifetimes, which have also been predicted by a recent \emph{ab
initio} calculation \cite{Spa01}.

\par

We thank W.\,Ekhardt and A.\,Rubio for helpful discussions and
M.\,Bovet for kindly providing the graphite band structure. It is
our pleasure to acknowledge G.\,Ertl for his continuing and
generous support. R.\,F.\,acknowledges financial support by the
Alexander von Humboldt foundation.



\begin{references}

\bibitem{Pet97} H.\,Petek and S.\,Ogawa,  Prog.\,Surf.\,Science {\bf 56}, 239 (1997).

\bibitem{Ech00} P.\,M.\,Echenique, J.\,M.\,Pitarke, E.\,V.\,Chulkov and A.\,Rubio,  Chem.\,Phys.\,{\bf 251}, 1 (2000).

\bibitem{Pin89} D.\,Pines, P.\,Nozieres, The theory of Quantum Liquids, vol.\,I: Normal Fermi Liquids, Addison-Wesley, New York, 1989.

\bibitem{Sch99} W.\,D.\,Sch\"one, R.\,Keyling, M.\,Bandic and W.\,Ekardt, Phys.\,Rev.\,B {\bf 60}, 8616 (1999).

\bibitem{Cam99} I.\,Campillo, J.\,M.\,Pitarke, A.\,Rubio, E.\,Zarate and P.\,M.\,Echenique , Phys.\,Rev.\,Lett., {\bf 83}, 2230 (1999).

\bibitem{Oga97} S.\,Ogawa, H.\,Nagano and H.\,Petek, Phys.\,Rev.\,B {\bf 55}, 10869 (1997).

\bibitem{Bau98} M.\,Bauer, S.\,Pawlik and M.\,Aeschlimann, Proc. \,SPIE. {\bf 3272}, 201 (1998).

\bibitem{Xu96} S.\,Xu, J.\,Cao, C.\,C.\,Miller, D.\,A.\,Mantell, R.\,J.\,D.\,Miller and Y.\,Gao,  Phys.\ Rev.\ Lett.\ {\bf 76}, 483 (1996).

\bibitem{Gon96} J.\ Gonzalez, F.\ Guinea and M.\ A.\ H.\ Vozmediano, Phys.\ Rev.\ Lett.\ , {\bf 77}, 3589 (1996).

\bibitem{Wal47} P.\,R.\,Wallace, Phys.\,Rev.\, {\bf 71}, 622 (1947).

\bibitem{Spa01} C.\ D.\ Spataru, M.\,A.\,Cazalilla, A.\,Rubio, L.X.\,Benedict, P.\,M.\,Echenique and Steven G.\,Louie, Phys.\ Rev.\ Lett.\ \emph{submitted}.

\bibitem{Kno98} E.\,Knoesel, A.\,Hotzel and M.\,Wolf, Phys.\,Rev.\,B {\bf 57}, 12812 (1998).

\bibitem{advcer} Advanced Ceramics Corporation, graphite monochromator plate, ZYB grade, $0.8\pm0.2°$ mosaic spread.

\bibitem{Ono76} S.\,Ono, J.\,Phys.\,Soc.\,Japan {\bf 40}, 498 (1976).

\bibitem{Her00a} T.\,Hertel and G.\,Moos, Phys.\,Rev.\,Lett., {\bf 84}, 5002 (2000).

\bibitem{Pet00} H.\,Petek, H.\,Nagano, M.\,J.\,Weida, S.\,Ogawa, Chem.\,Phys.\,, {\bf251}\, 71 (2000).

\bibitem{Moo01} G.\,Moos, \emph{et al.}, in preparation.

\bibitem{Her00} T.\,Hertel and G.\,Moos, Chem.\,Phys.\,Lett., {\bf 320}, 359 (2000).

\bibitem{Sei90} K.\,Seibert, G.\,C.\,Cho, W.\,K\"utt, H.\,Kurz, D.\,H.\,Reitze, J.\,I.\,Dadap, H.\,Ahn, M.\,C.\,Downer and A.\,M.\,Malvezzi, Phys.\,Rev.\,B {\bf 42}, 2842 (1990).

\bibitem{Wei99} M.\,Weinelt, C.\,Reu\ss, M.\,Kutschera, U.\,Thomann, I.\,L.\,Shumay, Th.\,Fauster, U.\,H\"ofer, F.\,Theilmann and A.\,Goldmann Appl.\,Phys.\,B, {\bf 68}, 377 (1999).

\bibitem{Ert99} K.\,Ertel, U.\,Kohl, J.\,Lehmann, M.\,Merschdorf, W.\,Pfeiffer, A.\,Thon, S.\,Voll and G.\,Gerber,  Appl.\,Phys.\,B {\bf 68}, 439 (1999).

\end{references}
\end{document}